\documentclass[graybox]{svmult}%
\usepackage{mathptmx}
\usepackage{helvet}
\usepackage{courier}
\usepackage{type1cm}
\usepackage{makeidx}
\usepackage{graphicx}
\usepackage{multicol}
\usepackage[bottom]{footmisc}
\usepackage{amsmath}
\usepackage{amsfonts}
\usepackage{amssymb}%
\makeindex
\begin{document}

\title{Gravitational Zero Point Energy and the Induced Cosmological Constant}
\author{Remo Garattini}

\institute{Remo Garattini \at Universit\`{a} degli Studi di Bergamo, Facolt\`{a} di Ingegneria,
\\ Viale Marconi 5, 24044 Dalmine (Bergamo) Italy and\\ I.N.F.N. -
sezione di Milano, Milan, Italy. \email{remo.garattini@unibg.it}}

\maketitle

\abstract{We discuss how to extract information about the cosmological constant from the
Wheeler-DeWitt equation, considered as an eigenvalue of a Sturm-Liouville
problem in a generic spherically symmetric background. The equation is
approximated to one loop with the help of a variational approach with Gaussian
trial wave functionals. A canonical decomposition of modes is used to separate
transverse-traceless tensors (graviton) from ghosts and scalar. We show that
no ghosts appear in the final evaluation of the cosmological constant. A zeta
function regularization and a ultra violet cutoff are used to handle with divergences.  A renormalization
procedure is introduced to remove the infinities. We compare the result with the one obtained in the
context of noncommutative geometries
}

\section{Introduction}

One of the biggest challenges of our century is the explanation of why the
observed cosmological constant is so small when compared to the one estimated
by Zero Point Energy (ZPE) computations in Quantum Field Theory. Indeed there
exists a difference of 120 orders of magnitude between them. However, it
appears that a definitive answer is still lacking. One possible approach to
this problem comes from the Wheeler-DeWitt equation (WDW)\cite{De Witt}, which
is described by%
\begin{equation}
\mathcal{H}\Psi=\left[  \left(  2\kappa\right)  G_{ijkl}\pi^{ij}\pi^{kl}%
-\frac{\sqrt{g}}{2\kappa}\!{}\!\left(  \,\!^{3}R-2\Lambda\right)  \right]
\Psi=0, \label{WDW1}%
\end{equation}
where $\kappa=8\pi G$, $G_{ijkl}$ is the super-metric and $^{3}R$ is the
scalar curvature in three dimensions. The main reason to use such an equation
is that its most general formulation intrinsically includes a cosmological
term. Moreover, if we formally re-write the WDW equation as\footnote{See also
Ref.\cite{CG:2007} for an application of the method to a $f\left(  R\right)  $
theory.}\cite{Remo}%
\begin{equation}
\frac{1}{V}\frac{\int\mathcal{D}\left[  g_{ij}\right]  \Psi^{\ast}\left[
g_{ij}\right]  \int_{\Sigma}d^{3}x\hat{\Lambda}_{\Sigma}\Psi\left[
g_{ij}\right]  }{\int\mathcal{D}\left[  g_{ij}\right]  \Psi^{\ast}\left[
g_{ij}\right]  \Psi\left[  g_{ij}\right]  }=\frac{1}{V}\frac{\left\langle
\Psi\left\vert \int_{\Sigma}d^{3}x\hat{\Lambda}_{\Sigma}\right\vert
\Psi\right\rangle }{\left\langle \Psi|\Psi\right\rangle }=-\frac{\Lambda
}{\kappa}, \label{WDW2}%
\end{equation}
where%
\begin{equation}
V=\int_{\Sigma}d^{3}x\sqrt{g}%
\end{equation}
is the volume of the hypersurface $\Sigma$ and%
\begin{equation}
\hat{\Lambda}_{\Sigma}=\left(  2\kappa\right)  G_{ijkl}\pi^{ij}\pi^{kl}%
-\sqrt{g}^{3}R/\left(  2\kappa\right)  ,
\end{equation}
we recognize that the WDW equation can be represented by an expectation value.
In particular, Eq.$\left(  \ref{WDW2}\right)  $ represents the Sturm-Liouville
problem associated with the cosmological constant. In this form the ratio
$\Lambda_{c}/\kappa$ represents the expectation value of $\hat{\Lambda
}_{\Sigma}$ without matter fields. The related boundary conditions are
dictated by the choice of the trial wave functionals which, in our case are of
the Gaussian type. Different types of wave functionals correspond to different
boundary conditions. The choice of a Gaussian wave functional is justified by
the fact that we would like to explain the cosmological constant $\left(
\Lambda_{c}/\kappa\right)  $ as a ZPE effect. To fix ideas, we will work with
the following form of the metric%
\begin{equation}
ds^{2}=-N^{2}\left(  r\right)  dt^{2}+\frac{dr^{2}}{1-\frac{b\left(  r\right)
}{r}}+r^{2}\left(  d\theta^{2}+\sin^{2}\theta d\phi^{2}\right)  , \label{dS}%
\end{equation}
where $b\left(  r\right)  $ is subject to the only condition $b\left(
r_{t}\right)  =r_{t}$. As a first step, we begin to decompose the
gravitational perturbation in such a way to obtain the graviton contribution
enclosed in Eq.$\left(  \ref{WDW2}\right)  $.

\section{Extracting the graviton contribution}

We can gain more information if we consider $g_{ij}=\bar{g}_{ij}+h_{ij},$where
$\bar{g}_{ij}$ is the background metric and $h_{ij}$ is a quantum fluctuation
around the background. Thus Eq.$\left(  \ref{WDW2}\right)  $ can be expanded
in terms of $h_{ij}$. Since the kinetic part of $\hat{\Lambda}_{\Sigma}$ is
quadratic in the momenta, we only need to expand the three-scalar curvature
$\int d^{3}x\sqrt{g}{}^{3}R$ up to the quadratic order. However, to proceed
with the computation, we also need an orthogonal decomposition on the tangent
space of 3-metric deformations\cite{Vassilevich:1993,Quad:1969}:%

\begin{equation}
h_{ij}=\frac{1}{3}\left(  \sigma+2\nabla\cdot\xi\right)  g_{ij}+\left(
L\xi\right)  _{ij}+h_{ij}^{\bot}.\label{p21a}%
\end{equation}
The operator $L$ maps $\xi_{i}$ into symmetric tracefree tensors%
\begin{equation}
\left(  L\xi\right)  _{ij}=\nabla_{i}\xi_{j}+\nabla_{j}\xi_{i}-\frac{2}%
{3}g_{ij}\left(  \nabla\cdot\xi\right)  ,
\end{equation}
$h_{ij}^{\bot}$ is the traceless-transverse component of the perturbation
(TT), namely $g^{ij}h_{ij}^{\bot}=0$, $\nabla^{i}h_{ij}^{\bot}=0$ and $h$ is
the trace of $h_{ij}$. It is immediate to recognize that the trace element
$\sigma=h-2\left(  \nabla\cdot\xi\right)  $ is gauge invariant. If we perform
the same decomposition also on the momentum $\pi^{ij}$, up to second order
Eq.$\left(  \ref{WDW2}\right)  $ becomes%
\begin{equation}
\frac{1}{V}\frac{\left\langle \Psi\left\vert \int_{\Sigma}d^{3}x\left[
\hat{\Lambda}_{\Sigma}^{\bot}+\hat{\Lambda}_{\Sigma}^{\xi}+\hat{\Lambda
}_{\Sigma}^{\sigma}\right]  ^{\left(  2\right)  }\right\vert \Psi\right\rangle
}{\left\langle \Psi|\Psi\right\rangle }=-\frac{\Lambda}{\kappa}%
.\label{lambda0_2}%
\end{equation}
Concerning the measure appearing in Eq.$\left(  \ref{WDW2}\right)  $, we have
to note that the decomposition $\left(  \ref{p21a}\right)  $ induces the
following transformation on the functional measure $\mathcal{D}h_{ij}%
\rightarrow\mathcal{D}h_{ij}^{\bot}\mathcal{D}\xi_{i}\mathcal{D}\sigma J_{1}$,
where the Jacobian related to the gauge vector variable $\xi_{i}$ is%
\begin{equation}
J_{1}=\left[  \det\left(  \bigtriangleup g^{ij}+\frac{1}{3}\nabla^{i}%
\nabla^{j}-R^{ij}\right)  \right]  ^{\frac{1}{2}}.
\end{equation}
This is nothing but the famous Faddev-Popov determinant. It becomes more
transparent if $\xi_{a}$ is further decomposed into a transverse part $\xi
_{a}^{T}$ with $\nabla^{a}\xi_{a}^{T}=0$ and a longitudinal part $\xi
_{a}^{\parallel}$ with $\xi_{a}^{\parallel}=$ $\nabla_{a}\psi$, then $J_{1}$
can be expressed by an upper triangular matrix for certain backgrounds (e.g.
Schwarzschild in three dimensions). It is immediate to recognize that for an
Einstein space in any dimension, cross terms vanish and $J_{1}$ can be
expressed by a block diagonal matrix. Since $\det AB=\det A\det B$, the
functional measure $\mathcal{D}h_{ij}$ factorizes into%
\begin{equation}
\mathcal{D}h_{ij}=\left(  \det\bigtriangleup_{V}^{T}\right)  ^{\frac{1}{2}%
}\left(  \det\left[  \frac{2}{3}\bigtriangleup^{2}+\nabla_{i}R^{ij}\nabla
_{j}\right]  \right)  ^{\frac{1}{2}}\mathcal{D}h_{ij}^{\bot}\mathcal{D}\xi
^{T}\mathcal{D}\psi
\end{equation}
with $\left(  \bigtriangleup_{V}^{ij}\right)  ^{T}=\bigtriangleup
g^{ij}-R^{ij}$ acting on transverse vectors, which is the Faddeev-Popov
determinant. In writing the functional measure $\mathcal{D}h_{ij}$, we have
here ignored the appearance of a multiplicative anomaly\cite{EVZ:1998}. Thus
the inner product can be written as%
\begin{equation}
\int\mathcal{D\rho}\Psi^{\ast}\left[  h_{ij}^{\bot}\right]  \Psi^{\ast}\left[
\xi^{T}\right]  \Psi^{\ast}\left[  \sigma\right]  \Psi\left[  h_{ij}^{\bot
}\right]  \Psi\left[  \xi^{T}\right]  \Psi\left[  \sigma\right]  ,
\end{equation}
where%
\begin{equation}
\mathcal{D}\rho=\mathcal{D}h_{ij}^{\bot}\mathcal{D}\xi^{T}\mathcal{D}%
\sigma\left(  \det\bigtriangleup_{V}^{T}\right)  ^{\frac{1}{2}}\left(
\det\left[  \frac{2}{3}\bigtriangleup^{2}+\nabla_{i}R^{ij}\nabla_{j}\right]
\right)  ^{\frac{1}{2}}.
\end{equation}
Nevertheless, since there is no interaction between ghost fields and the other
components of the perturbation at this level of approximation, the Jacobian
appearing in the numerator and in the denominator simplify. The reason can be
found in terms of connected and disconnected terms. The disconnected terms
appear in the Faddeev-Popov determinant and these ones are not linked by the
Gaussian integration. This means that disconnected terms in the numerator and
the same ones appearing in the denominator cancel out. Therefore, Eq.$\left(
\ref{lambda0_2}\right)  $ factorizes into three pieces. The piece containing
$\hat{\Lambda}_{\Sigma}^{\bot}$ is the contribution of the
transverse-traceless tensors (TT): essentially is the graviton contribution
representing true physical degrees of freedom. Regarding the vector term
$\hat{\Lambda}_{\Sigma}^{T}$, we observe that under the action of
infinitesimal diffeomorphism generated by a vector field $\epsilon_{i}$, the
components of $\left(  \ref{p21a}\right)  $ transform as
follows\cite{Vassilevich:1993}%
\begin{equation}
\xi_{j}\longrightarrow\xi_{j}+\epsilon_{j},\qquad h\longrightarrow
h+2\nabla\cdot\xi,\qquad h_{ij}^{\bot}\longrightarrow h_{ij}^{\bot}.
\end{equation}
The Killing vectors satisfying the condition $\nabla_{i}\xi_{j}+\nabla_{j}%
\xi_{i}=0,$ do not change $h_{ij}$, and thus should be excluded from the gauge
group. All other diffeomorphisms act on $h_{ij}$ nontrivially. We need to fix
the residual gauge freedom on the vector $\xi_{i}$. The simplest choice is
$\xi_{i}=0$. This new gauge fixing produces the same Faddeev-Popov determinant
connected to the Jacobian $J_{1}$ and therefore will not contribute to the
final value. We are left with%
\begin{equation}
\frac{1}{V}\frac{\left\langle \Psi^{\bot}\left\vert \int_{\Sigma}d^{3}x\left[
\hat{\Lambda}_{\Sigma}^{\bot}\right]  ^{\left(  2\right)  }\right\vert
\Psi^{\bot}\right\rangle }{\left\langle \Psi^{\bot}|\Psi^{\bot}\right\rangle
}+\frac{1}{V}\frac{\left\langle \Psi^{\sigma}\left\vert \int_{\Sigma}%
d^{3}x\left[  \hat{\Lambda}_{\Sigma}^{\sigma}\right]  ^{\left(  2\right)
}\right\vert \Psi^{\sigma}\right\rangle }{\left\langle \Psi^{\sigma}%
|\Psi^{\sigma}\right\rangle }=-\frac{\Lambda^{\bot}}{\kappa}-\frac
{\Lambda^{\sigma}}{\kappa}.\label{lambda0_2a}%
\end{equation}
Note that in the expansion of $\int_{\Sigma}d^{3}x\sqrt{g}{}R$ to second
order, a coupling term between the TT component and scalar one remains.
However, the Gaussian integration does not allow such a mixing which has to be
introduced with an appropriate wave functional. Extracting the TT tensor
contribution from Eq.$\left(  \ref{WDW2}\right)  $ approximated to second
order in perturbation of the spatial part of the metric into a background term
$\bar{g}_{ij}$, and a perturbation $h_{ij}$, we get%
\begin{equation}
\hat{\Lambda}_{\Sigma}^{\bot}=\frac{1}{4V}\int_{\Sigma}d^{3}x\sqrt{\bar{g}%
}G^{ijkl}\left[  \left(  2\kappa\right)  K^{-1\bot}\left(  x,x\right)
_{ijkl}+\frac{1}{\left(  2\kappa\right)  }\!{}\left(  \tilde{\bigtriangleup
}_{L\!}\right)  _{j}^{a}K^{\bot}\left(  x,x\right)  _{iakl}\right]
,\label{p22}%
\end{equation}
where%
\begin{equation}
\left(  \tilde{\bigtriangleup}_{L\!}\!{}h^{\bot}\right)  _{ij}=\left(
\bigtriangleup_{L\!}\!{}h^{\bot}\right)  _{ij}-4R{}_{i}^{k}\!{}h_{kj}^{\bot
}+\text{ }^{3}R{}\!{}h_{ij}^{\bot}\label{M Lichn}%
\end{equation}
is the modified Lichnerowicz operator and $\bigtriangleup_{L}$is the
Lichnerowicz operator defined by%
\begin{equation}
\left(  \bigtriangleup_{L}h\right)  _{ij}=\bigtriangleup h_{ij}-2R_{ikjl}%
h^{kl}+R_{ik}h_{j}^{k}+R_{jk}h_{i}^{k}\qquad\bigtriangleup=-\nabla^{a}%
\nabla_{a}.\label{DeltaL}%
\end{equation}
$G^{ijkl}$ represents the inverse DeWitt metric and all indices run from one
to three. Note that the term $-4R{}_{i}^{k}\!{}h_{kj}^{\bot}+$ $^{3}R{}%
\!{}h_{ij}^{\bot}$ disappears in four dimensions. The propagator $K^{\bot
}\left(  x,x\right)  _{iakl}$ can be represented as
\begin{equation}
K^{\bot}\left(  \overrightarrow{x},\overrightarrow{y}\right)  _{iakl}%
=\sum_{\tau}\frac{h_{ia}^{\left(  \tau\right)  \bot}\left(  \overrightarrow
{x}\right)  h_{kl}^{\left(  \tau\right)  \bot}\left(  \overrightarrow
{y}\right)  }{2\lambda\left(  \tau\right)  },\label{proptt}%
\end{equation}
where $h_{ia}^{\left(  \tau\right)  \bot}\left(  \overrightarrow{x}\right)  $
are the eigenfunctions of $\tilde{\bigtriangleup}_{L\!}$. $\tau$ denotes a
complete set of indices and $\lambda\left(  \tau\right)  $ are a set of
variational parameters to be determined by the minimization of Eq.$\left(
\ref{p22}\right)  $. The expectation value of $\hat{\Lambda}_{\Sigma}^{\bot}$
is easily obtained by inserting the form of the propagator into Eq.$\left(
\ref{p22}\right)  $ and minimizing with respect to the variational function
$\lambda\left(  \tau\right)  $. Thus the total one loop energy density for TT
tensors becomes%
\begin{equation}
\frac{\Lambda}{8\pi G}=-\frac{1}{2}\sum_{\tau}\left[  \sqrt{\omega_{1}%
^{2}\left(  \tau\right)  }+\sqrt{\omega_{2}^{2}\left(  \tau\right)  }\right]
.\label{1loop}%
\end{equation}
The above expression makes sense only for $\omega_{i}^{2}\left(  \tau\right)
>0$, where $\omega_{i}$ are the eigenvalues of $\tilde{\bigtriangleup}_{L\!}$.
In the next section, we will explicitly evaluate Eq.$\left(  \ref{1loop}%
\right)  $ for a background of spherically symmetric type.

\section{One loop energy density}

\subsection{ Conventional Regularization and Renormalization}

The reference metric $\left(  \ref{dS}\right)  $ can be cast into the
following form%
\begin{equation}
ds^{2}=-N^{2}\left(  r\left(  x\right)  \right)  dt^{2}+dx^{2}+r^{2}\left(
x\right)  \left(  d\theta^{2}+\sin^{2}\theta d\phi^{2}\right)  ,\label{metric}%
\end{equation}
where%
\begin{equation}
dx=\pm\frac{dr}{\sqrt{1-\frac{b\left(  r\right)  }{r}}}\label{dx}%
\end{equation}
and $b\left(  r\right)  $ a generic shape function. Specific examples are%
\begin{equation}
b\left(  r\right)  =\frac{\Lambda_{dS}}{3}r^{3};\qquad b\left(  r\right)
=-\frac{\Lambda_{AdS}}{3}r^{3}\qquad\mathrm{and}\qquad b\left(  r\right)
=2MG.
\end{equation}
However, we would like to maintain the form of the line element $\left(
\ref{metric}\right)  $ as general as possible. With the help of Regge and
Wheeler representation\cite{Regge Wheeler:1957}, the Lichnerowicz operator
$\left(  \tilde{\bigtriangleup}_{L\!}\!{}h^{\bot}\right)  _{ij}$ can be
reduced to%
\begin{equation}
\left[  -\frac{d^{2}}{dx^{2}}+\frac{l\left(  l+1\right)  }{r^{2}}+m_{i}%
^{2}\left(  r\right)  \right]  f_{i}\left(  x\right)  =\omega_{i,l}^{2}%
f_{i}\left(  x\right)  \quad i=1,2\quad,\label{p34}%
\end{equation}
where we have used reduced fields of the form $f_{i}\left(  x\right)
=F_{i}\left(  x\right)  /r$ and where we have defined two r-dependent
effective masses $m_{1}^{2}\left(  r\right)  $ and $m_{2}^{2}\left(  r\right)
$%
\begin{equation}
\left\{
\begin{array}
[c]{c}%
m_{1}^{2}\left(  r\right)  =\frac{6}{r^{2}}\left(  1-\frac{b\left(  r\right)
}{r}\right)  +\frac{3}{2r^{2}}b^{\prime}\left(  r\right)  -\frac{3}{2r^{3}%
}b\left(  r\right)  \\
\\
m_{2}^{2}\left(  r\right)  =\frac{6}{r^{2}}\left(  1-\frac{b\left(  r\right)
}{r}\right)  +\frac{1}{2r^{2}}b^{\prime}\left(  r\right)  +\frac{3}{2r^{3}%
}b\left(  r\right)
\end{array}
\right.  \quad\left(  r\equiv r\left(  x\right)  \right)  .\label{masses}%
\end{equation}
In order to use the W.K.B. method considered by `t Hooft in the brick wall
problem\cite{tHooft:1985}, from Eq.$\left(  \ref{p34}\right)  $ we can extract
two r-dependent radial wave numbers%
\begin{equation}
k_{i}^{2}\left(  r,l,\omega_{i,nl}\right)  =\omega_{i,nl}^{2}-\frac{l\left(
l+1\right)  }{r^{2}}-m_{i}^{2}\left(  r\right)  \quad i=1,2\quad.\label{kTT}%
\end{equation}
Then the counting of the number of modes with frequency less than $\omega_{i}$
is given approximately by%
\begin{equation}
\tilde{g}\left(  \omega_{i}\right)  =\int_{0}^{l_{\max}}\nu_{i}\left(
l,\omega_{i}\right)  \left(  2l+1\right)  dl.\label{p41}%
\end{equation}
$\nu_{i}\left(  l,\omega_{i}\right)  $ is the number of nodes in the mode with
$\left(  l,\omega_{i}\right)  $, such that $\left(  r\equiv r\left(  x\right)
\right)  $
\begin{equation}
\nu_{i}\left(  l,\omega_{i}\right)  =\frac{1}{\pi}\int_{-\infty}^{+\infty
}dx\sqrt{k_{i}^{2}\left(  r,l,\omega_{i}\right)  }.\label{p42}%
\end{equation}
Here it is understood that the integration with respect to $x$ and $l_{\max}$
is taken over those values which satisfy $k_{i}^{2}\left(  r,l,\omega
_{i}\right)  \geq0$. With the help of Eqs.$\left(  \ref{p41},\ref{p42}\right)
$, Eq.$\left(  \ref{1loop}\right)  $ becomes%
\begin{equation}
\frac{\Lambda}{8\pi G}=-\frac{1}{\pi}\sum_{i=1}^{2}\int_{0}^{+\infty}%
\omega_{i}\frac{d\tilde{g}\left(  \omega_{i}\right)  }{d\omega_{i}}d\omega
_{i}.\label{tot1loop}%
\end{equation}
This is the one loop graviton contribution to the induced cosmological
constant. The explicit evaluation of Eq.$\left(  \ref{tot1loop}\right)  $
gives%
\begin{equation}
\frac{\Lambda}{8\pi G}=\rho_{1}+\rho_{2}=-\frac{1}{4\pi^{2}}\sum_{i=1}^{2}%
\int_{\sqrt{m_{i}^{2}\left(  r\right)  }}^{+\infty}\omega_{i}^{2}\sqrt
{\omega_{i}^{2}-m_{i}^{2}\left(  r\right)  }d\omega_{i},\label{t1l}%
\end{equation}
where we have included an additional $4\pi$ coming from the angular
integration. The use of the zeta function regularization method to compute the
energy densities $\rho_{1}$ and $\rho_{2}$ leads to%
\begin{equation}
\rho_{i}\left(  \varepsilon\right)  =\frac{m_{i}^{4}\left(  r\right)  }%
{64\pi^{2}}\left[  \frac{1}{\varepsilon}+\ln\left(  \frac{4\mu^{2}}{m_{i}%
^{2}\left(  r\right)  \sqrt{e}}\right)  \right]  \quad i=1,2\quad,\label{rhoe}%
\end{equation}
where we have introduced the additional mass parameter $\mu$ in order to
restore the correct dimension for the regularized quantities. Such an
arbitrary mass scale emerges unavoidably in any regularization scheme. The
renormalization is performed via the absorption of the divergent part into the
re-definition of a bare classical quantity. Here we have two possible choices:
the induced cosmological constant $\Lambda$ or the gravitational Newton
constant $G$. In any case a certain degree of arbitrariness is present because
of the scale parameter $\mu$. However, it is instructive a comparison of the
result in Eq.$\left(  \ref{rhoe}\right)  $ with the one which can be obtained
by imposing a UV cutoff. A direct calculation leads to $\left(  i=1,2\right)
$%
\[
\int_{\sqrt{m_{i}^{2}\left(  r\right)  }}^{+\infty}\omega_{i}^{2}\sqrt
{\omega_{i}^{2}-m_{i}^{2}\left(  r\right)  }d\omega_{i}%
\]%
\[
\underset{x_{i}=\omega_{i}/\sqrt{m_{i}^{2}\left(  r\right)  }}{=}\frac
{m_{i}^{4}\left(  r\right)  }{4}\left[  x_{i}^{3}\sqrt{x_{i}^{2}-1}%
-\frac{x_{i}}{2}\sqrt{x_{i}^{2}-1}-\frac{1}{2}\ln\left(  x_{i}+\sqrt{x_{i}%
^{2}-1}\right)  \right]  _{1}^{\omega_{UV}/\sqrt{m_{i}^{2}\left(  r\right)  }}%
\]%
\begin{equation}
\simeq\frac{m_{i}^{4}\left(  r\right)  }{4}\left[  \frac{\omega_{UV}^{4}%
}{m_{i}^{4}\left(  r\right)  }-\frac{\omega_{UV}^{2}}{2m_{i}^{2}\left(
r\right)  }-\frac{1}{2}\ln\left(  \frac{2\omega_{UV}}{\sqrt{m_{i}^{2}\left(
r\right)  }}\right)  \right]  ,\label{UVm}%
\end{equation}
where $\omega_{UV}\gg\sqrt{m_{i}^{2}\left(  r\right)  }$. Nevertheless, for
some backgrounds in some ranges,
\begin{equation}
m_{0}^{2}\left(  r\right)  =m_{1}^{2}\left(  r\right)  =-m_{2}^{2}\left(
r\right)  .\label{masses1}%
\end{equation}
Thus, in these cases%
\[
\frac{\Lambda}{8\pi G}=\rho_{1}+\rho_{2}=-\frac{1}{4\pi^{2}}\left[
\int_{\sqrt{m_{0}^{2}\left(  r\right)  }}^{+\infty}\omega^{2}\sqrt{\omega
^{2}-m_{0}^{2}\left(  r\right)  }d\omega+\int_{0}^{+\infty}\omega^{2}%
\sqrt{\omega^{2}+m_{0}^{2}\left(  r\right)  }d\omega\right]
\]%
\begin{equation}
\simeq-\frac{1}{4\pi^{2}}\left[  \frac{\omega_{UV}^{4}}{2}+\frac{m_{0}%
^{4}\left(  r\right)  }{8}\ln\left(  \frac{m_{0}^{2}\left(  r\right)  \sqrt
{e}}{4\omega_{UV}^{2}}\right)  \right]  ,
\end{equation}
where we have used%
\[
\int_{0}^{+\infty}\omega^{2}\sqrt{\omega^{2}+m_{0}^{2}\left(  r\right)
}d\omega
\]%
\begin{equation}
\underset{x=\omega/\sqrt{m_{0}^{2}\left(  r\right)  }}{=}\frac{m_{0}%
^{4}\left(  r\right)  }{4}\left[  x^{3}\sqrt{x^{2}+1}+\frac{x}{2}\sqrt
{x^{2}+1}-\frac{1}{2}\ln\left(  x+\sqrt{x^{2}+1}\right)  \right]  _{0}%
^{\omega_{UV}/\sqrt{m_{0}^{2}\left(  r\right)  }}.
\end{equation}
The Schwarzschild Schwarzschild-de Sitter (SdS) and Schwarzschild-Anti de
Sitter (SAdS) backgrounds satisfy relation $\left(  \ref{masses1}\right)  $ in
a region close to the throat. Indeed, by expanding $b\left(  r\right)  $ close
to the throat, one gets $\left(  r\equiv r\left(  x\right)  \right)  $%
\begin{equation}
\left\{
\begin{array}
[c]{c}%
m_{1}^{2}\left(  r\right)  =\frac{6}{r^{2}}-\frac{15r_{t}}{2r^{3}}%
-\frac{6b^{\prime}\left(  r_{t}\right)  }{r^{2}}+\frac{15b^{\prime}\left(
r_{t}\right)  r_{t}}{2r^{3}}\\
\\
m_{2}^{2}\left(  r\right)  =\frac{6}{r^{2}}-\frac{9r_{t}}{2r^{3}}%
-\frac{4b^{\prime}\left(  r_{t}\right)  }{r^{2}}+\frac{9b^{\prime}\left(
r_{t}\right)  r_{t}}{2r^{3}}%
\end{array}
\right.
\end{equation}
and for example, for the Schwarzschild case where $b\left(  r\right)
=r_{t}=2MG$, we get
\begin{equation}
\qquad\left\{
\begin{array}
[c]{c}%
m_{1}^{2}\left(  r\right)  =-\frac{3r_{t}}{2r^{3}}\\
\\
m_{2}^{2}\left(  r\right)  =+\frac{3r_{t}}{2r^{3}}%
\end{array}
\right.  .
\end{equation}
Note that Eq.$\left(  \ref{lambda0}\right)  $ works when the effective masses
satisfy relation $\left(  \ref{masses1}\right)  $, otherwise the zeta function
and the cutoff regularizations produce different results as shown by
Eq.$\left(  \ref{UVm}\right)  $. The divergence can be eliminated by
separating the cosmological constant $\Lambda$, into a bare cosmological
constant $\Lambda_{0}$ and a divergent quantity $\Lambda^{div}$, where%
\begin{equation}
\Lambda^{div}=\frac{Gm_{0}^{4}\left(  r\right)  }{\varepsilon32\pi^{2}},
\end{equation}
or%
\begin{equation}
\Lambda_{UV}^{div}=-\frac{G}{4\pi^{2}}\left[  \frac{\omega_{UV}^{4}}{2}%
+\frac{m_{0}^{4}\left(  r\right)  }{8}\ln\left(  \frac{\mu^{2}\sqrt{e}%
}{4\omega_{UV}^{2}}\right)  \right]  .
\end{equation}
In both cases, the remaining finite value for the cosmological constant reads%
\begin{equation}
\frac{\Lambda_{0}}{8\pi G}=\left(  \rho_{1}\left(  \mu\right)  +\rho
_{2}\left(  \mu\right)  \right)  =\rho_{eff}^{TT}\left(  \mu,r\right)
=\frac{m_{0}^{4}\left(  r\right)  }{32\pi^{2}}\ln\left(  \frac{4\mu^{2}}%
{m_{0}^{2}\left(  r\right)  \sqrt{e}}\right)  .\label{lambda0}%
\end{equation}

\subsection{The example of Non Commutative theories}

Non Commutative theories provide a powerful method to naturally regularize
divergent integrals appearing in Eq.$\left(  \ref{t1l}\right)  $. Basically,
the number of states is modified in the following way\cite{RG PN}%
\begin{equation}
dn=\frac{d^{3}xd^{3}k}{\left(  2\pi\right)  ^{3}}\ \Longrightarrow
\ dn_{i}=\frac{d^{3}xd^{3}k}{\left(  2\pi\right)  ^{3}}\exp\left(
-\frac{\theta}{4}k_{i}^{2}\right)  , \label{moddn}%
\end{equation}
with%
\begin{equation}
k_{i}^{2}=\omega_{i,nl}^{2}-m_{i}^{2}\left(  r\right)  \quad i=1,2.
\end{equation}
This deformation corresponds to an effective cut off on the background
geometry $\left(  \ref{metric}\right)  $. The UV cut off is triggered only by
higher momenta modes $\gtrsim1/\sqrt{\theta}$ which propagate over the
background geometry. The virtue of this kind of deformation is its exponential
damping profile, which encodes an intrinsic nonlocal character into fields
$f_{i}(x)$. Plugging $\left(  \ref{p42}\right)  $ into $\left(  \ref{p41}%
\right)  $ and taking account of $\left(  \ref{moddn}\right)  $, the number of
modes with frequency less than $\omega_{i}$, $i=1,2$ is given by%
\begin{equation}
\tilde{g}\left(  \omega_{i}\right)  =\frac{1}{\pi}\int_{-\infty}^{+\infty
}dx\int_{0}^{l_{\max}}\left(  2l+1\right)  \sqrt{\omega_{i,nl}^{2}%
-\frac{l\left(  l+1\right)  }{r^{2}}-m_{i}^{2}\left(  r\right)  }\exp\left(
-\frac{\theta}{4}k_{i}^{2}\right)  \ dl
\end{equation}
and the induced cosmological constant becomes%
\begin{equation}
\frac{\Lambda}{8\pi G}=\frac{1}{6\pi^{2}}\left[  \int_{\sqrt{m_{0}^{2}\left(
r\right)  }}^{+\infty}\sqrt{\left(  \omega^{2}-m_{0}^{2}\left(  r\right)
\right)  ^{3}}e^{-\frac{\theta}{4}\left(  \omega^{2}-m_{0}^{2}\left(
r\right)  \right)  }+\int_{0}^{+\infty}\sqrt{\left(  \omega^{2}+m_{0}%
^{2}\left(  r\right)  \right)  ^{3}}e^{-\frac{\theta}{4}\left(  \omega
^{2}+m_{0}^{2}\left(  r\right)  \right)  }\right]  , \label{t1loop}%
\end{equation}
which integrated leads to%
\begin{equation}
\frac{\Lambda}{8\pi G}=\frac{1}{12\pi^{2}}\left(  \frac{4}{\theta}\right)
^{2}\left(  y\cosh\left(  \frac{y}{2}\right)  -y^{2}\sinh\left(  \frac{y}%
{2}\right)  \right)  \ K_{1}\left(  \frac{y}{2}\right)  +y^{2}\cosh\left(
\frac{y}{2}\right)  K_{0}\left(  \frac{y}{2}\right)  , \label{LambdaNCS}%
\end{equation}
where $K_{0}\left(  y\right)  $ and $K_{1}\left(  y\right)  $ are the modified
Bessel function and%
\begin{equation}
y=\frac{m_{0}^{2}\left(  r\right)  \theta}{4}. \label{xS}%
\end{equation}
The asymptotic properties of $\left(  \ref{LambdaNCS}\right)  $ show that the
one loop contribution is everywhere regular. Indeed, we find that when
$y\rightarrow+\infty$,
\begin{equation}
\frac{\Lambda}{8\pi G}\simeq\frac{1}{6\pi^{2}\theta^{2}}\sqrt{\frac{\pi}{y}%
}\left[  3+\left(  8y^{2}+6y+3\right)  \exp\left(  -y\right)  \right]
\rightarrow0. \label{LNCSz}%
\end{equation}
Conversely, when $y\rightarrow0$, we obtain%
\begin{equation}
\frac{\Lambda}{8\pi G}\simeq\frac{4}{3\pi^{2}\theta^{2}}\left[  2-\left(
\frac{7}{8}+\frac{3}{4}\ln\left(  \frac{y}{4}\right)  +\frac{3}{4}%
\gamma\right)  y^{2}\right]  \rightarrow\frac{8}{3\pi^{2}\theta^{2}}%
\end{equation}
a finite value for $\Lambda$. Note that expression $\left(  \ref{LambdaNCS}%
\right)  $ can be used when the background satisfies the relation $\left(
\ref{masses1}\right)  $. For the other cases, we find that the effective
masses contribute in the same way at one loop. Thus $\left(  \ref{t1loop}%
\right)  $ becomes
\begin{equation}
\frac{\Lambda}{8\pi G}=\frac{1}{6\pi^{2}}\left[  \int_{\sqrt{m_{1}^{2}\left(
r\right)  }}^{+\infty}\sqrt{\left(  \omega^{2}-m_{1}^{2}\left(  r\right)
\right)  ^{3}}e^{-\frac{\theta}{4}\left(  \omega^{2}-m_{1}^{2}\left(
r\right)  \right)  }+\int_{\sqrt{m_{2}^{2}\left(  r\right)  }}^{+\infty}%
\sqrt{\left(  \omega^{2}-m_{2}^{2}\left(  r\right)  \right)  ^{3}}%
e^{-\frac{\theta}{4}\left(  \omega^{2}-m_{2}^{2}\left(  r\right)  \right)
}\right]  . \label{t2l}%
\end{equation}
For example, when%
\begin{equation}
m_{1}^{2}\left(  r\right)  =m_{2}^{2}\left(  r\right)  ,
\end{equation}
Eq.$\left(  \ref{t2l}\right)  $ reduces to%
\begin{equation}
\frac{\Lambda}{8\pi G}=\frac{1}{6\pi^{2}}\left(  \frac{4}{\theta}\right)
^{2}\left(  \frac{1}{2}y\left(  1-y\right)  K_{1}\left(  \frac{y}{2}\right)
+\frac{1}{2}y^{2}K_{0}\left(  \frac{y}{2}\right)  \right)  \exp\left(
\frac{y}{2}\right)  . \label{LNCdSAdS}%
\end{equation}
The asymptotic expansion of Eq.$\left(  \ref{LNCdSAdS}\right)  $ leads to%
\begin{equation}
\frac{\Lambda}{8\pi G}\simeq\frac{1}{6\pi^{2}}\left(  \frac{4}{\theta}\right)
^{2}\frac{3}{8}\sqrt{\frac{\pi}{y}}\rightarrow0,
\end{equation}
when $y\rightarrow\infty$. On the other hand, when $z\rightarrow0$, one gets%
\begin{equation}
\frac{\Lambda}{8\pi G}\simeq\frac{1}{6\pi^{2}}\left(  \frac{4}{\theta}\right)
^{2}\left[  1-\frac{z}{2}+\left(  -\frac{7}{16}-\frac{3}{8}\ln\left(  \frac
{z}{4}\right)  -\frac{3}{8}\gamma\right)  z^{2}\right]  \rightarrow\frac
{8}{3\pi^{2}\theta^{2}},
\end{equation}
i.e. a finite value of the cosmological term.

\section{Summary and Conclusions}

In this contribution, the effect of a ZPE on the cosmological constant has
been investigated using two specific geometries such as dS and AdS metrics.
The computation has been done by means of a variational procedure with a
Gaussian Wave Functional which should be a good candidate for a ZPE
calculation. We have found that only the graviton is
relevant\cite{GriKos:1989}. Actually, the appearance of a ghost contribution
is connected with perturbations of the shift vectors\cite{Vassilevich:1993}.
In this work we have excluded such perturbations. As usual, in ZPE calculation
we meet the problem of divergences which are regularized with zeta function
techniquesor by introducting a UV cutoff. After regularization , we have
adopted to remove divergences by absorbing them into the induced cosmological
constant $\Lambda$. Another possibility of keeping under control divergences
comes from a NCG induced minimal length. As a result we get a modified
counting of graviton modes. This let us obtain everywhere regular values for
the cosmological constant, independently of the chosen background, which
nevertheless is of a spherically symmetric type. Although the result seems to
be promising, we have to note that the evaluation is at the Planck scale.

\end{document}